\newcommand{\BSCCO}{$\rm{Bi_2Sr_2CaCu_2O_{8+\delta}}$ }
\newcommand{\LCO}{$\rm{La_2CuO_{4+\it{x}}}$ }
\newcommand{\LSCO}{$\rm{La_{\it{x}}Sr_{1-\it{x}}CuO_4}$ }
\newcommand{\LNSCO}{$\rm{La_{1.6-\it{x}}Nd_{0.4}Sr_{\it{x}} CuO_4}$ }
\newcommand{\YBCO}{$\rm{YBa_2Cu_3O_{6.35}}$ }

\documentstyle[aps,prl,epsf,floats]{revtex}
\begin{document}
\twocolumn[\hsize\textwidth\columnwidth\hsize\csname@twocolumnfalse%
\endcsname

\title{Translational Symmetry Breaking in the Superconducting State
of the Cuprates: 
\newline
Analysis of the Quasiparticle Density of States}
\author{Daniel Podolsky, Eugene Demler, Kedar Damle, and B.I. Halperin}
\address{Department of Physics, Harvard University, Cambridge MA 02138}

\maketitle

\begin{abstract}
Motivated by recent STM experiments on \BSCCO
\cite{davis1,kapitulnik1,davis2,davis3,kapitulnik2}, we study
the effects of weak translational symmetry breaking on the
quasiparticle spectrum of a $d$-wave superconductor.  We develop
a general formalism to discuss periodic charge
order, as well as quasiparticle scattering off localized defects.
We argue that the STM experiments in
\BSCCO cannot be explained using a simple charge density wave order
parameter, but are consistent with the presence of a periodic
modulation in the electron hopping or pairing amplitude. We review the
effects of randomness and pinning of the charge order and compare it
to the impurity scattering of quasiparticles.  We also discuss
implications of weak translational symmetry breaking for ARPES
experiments.
\end{abstract}
 ]

\section{Introduction}
Puzzling properties of the  high-$T_c$ cuprates have often been attributed to
the existence of competing instabilities, and 
proximity (or even coexistence) of several ordered states.
Possible instabilities that have been discussed in this context
include charge density wave (CDW) order, non-two-sublattice spin density
wave (SDW) order~\cite{zaanen,tranquada99,zhang,polkovnikov}, spin Peierls
order~\cite{vojta}, and orbital magnetism~\cite{chakravarty,varma}.
Neutron scattering
experiments on  \LNSCO~\cite{tranquada97},
\LSCO~\cite{katano,lake}, and
\LCO ~\cite{khaykovich} demonstrated the coexistence
of magnetism and superconductivity, while
recent experiments on strongly underdoped \YBCO \cite{mook}
have seen evidence of CDW order coexisting with superconductivity.
Particularly striking in this context are recent
STM experiments on \BSCCO \cite{davis1,kapitulnik1,kapitulnik2}
which see spatial structure in the tunneling density of states with a
period of four lattice constants.
This structure was originally observed
in the experiments in a magnetic field by 
J.E. Hoffman {\it et al.}
\cite{davis1}, and later also seen in zero field
by C. Howald {\it et al.} \cite{kapitulnik1,kapitulnik2}.
Modulo certain experimental subtleties,
these experiments can be thought of as measurements of the
spatial Fourier component (at the ordering wavevector ${\bf Q}$)
of the energy dependent local density
of states (LDOS) $\rho_{\bf Q}(\epsilon)$.

In this paper, we demonstrate that the energy dependence
of $\rho_{\bf Q}(\epsilon)$ provides important information
about the nature of charge ordering
in these materials. It allows us to separate simple charge
density wave  order,
that has only the Hartree-Fock potential modulation,
from the more unusual charge orders that involve modulation of the
electron kinetic energy (dimerization) or
the pairing amplitude (anomalous dimerization).  
For example, when ${\bf Q}=(2\pi/4,0)$ (as observed in slightly 
overdoped \BSCCO),  
there is a change of sign in $\rho_{\bf Q}(\epsilon)$ for energies 
around 40 meV when dimerizations are present, but not in the case of a 
simple CDW. 
When several of the simple distortions are present simultaneously, we can 
understand the resulting $\rho_{\bf Q}(\epsilon)$ as (roughly) a superposition of the 
corresponding simple cases, since the induced 
$\rho_{\bf Q}(\epsilon)$ is 
approximately linear in the 
order parameter for small distortions. Such a superposition is necessary to 
understand the experiments of \cite{kapitulnik1,kapitulnik2}.  

This superposition principle also applies when we have potential 
modulation at several
wavevectors, and $\rho_{\bf q}(\epsilon)$ can be analyzed separately
for each wavevector ${\bf q}$. This is necessary, for example, when we have
randomness that pins the charge order, so that the 
single particle potential is
not a delta function in momentum space but has a narrow distribution
centered at the ordering wavevector ${\bf Q}$. This leads to a finite
$\rho_{\bf q}(\epsilon)$ for a range
of wavevectors around ${\bf Q}$ and, as we
discuss below, taking a reasonable value of the CDW correlation length
reproduces well the ``weak dispersion'' of the
CDW peak observed in \cite{kapitulnik2}.  Our analysis can be extended
to systems with no charge order but, instead, with localized defects,
e.g. impurities. In this case we have a potential that is essentially
momentum independent and we find strongly dispersing peaks in
$\rho_{\bf q}(\epsilon)$ for a wide range of wavevectors. Such peaks
have been observed in the STM experiments in \cite{davis2,davis3}
and discussed theoretically in \cite{dhlee,polkovnikov2}.
We provide a qualitative comparison of the
STM spectra for systems with disordered CDW and impurity
scattering.

It is common to discuss spin density wave (SDW) order
as the primary competitor to superconductivity in the
underdoped cuprates \cite{zaanen,zhang,schulz,pryadko,yzhang}.
An order parameter for non-two-sublattice magnetism is
\begin{eqnarray}
\vec{S}({\bf r}) = \vec{\phi} e^{i {\bf Q}_s {\bf r}}
+ \vec{\phi}^* e^{-i {\bf Q}_s {\bf r}},
\label{SDW}
\end{eqnarray}
where the complex-valued vector $\vec{\phi}$
acquires an expectation value in a state with broken spin symmetry.
When the SDW order in (\ref{SDW}) is collinear, it has 
an associated spin singlet order parameter
that only breaks translational symmetry and can be
described as a generalized charge density wave \cite{yzhang}
\begin{eqnarray}
\delta \rho({\bf r}) = \varphi e^{i {\bf Q}_c {\bf r}}
+ \varphi^{*} e^{-i {\bf Q}_c {\bf r}}.
\label{gCDW}
\end{eqnarray}
Symmetry arguments determine the wavevector
${\bf Q}_c = 2 {\bf Q}_s $ of such generalized 
CDW, but they do not clarify its internal
structure. For example, modulation of the
local Hartree-Fock potential of the electrons and modulation
of the electron kinetic energy (hopping) are both spin singlet order
parameters that can be defined at the same wavevector and
described by (\ref{gCDW}). Modulation of the electron
pairing amplitude also belongs to the same
class of translational symmetry breaking since, in the
superconducting state with condensed Cooper pairs,
order parameters with charge two and zero are not orthogonal.
It is important to note, however, that a long range SDW order
is not a prerequisite for translational
symmetry breaking. One can have a situation where
quantum or thermal fluctuations destroy the spin order
but preserve a long range order in the charge
sector \cite{EK}. This was observed, for example, 
in underdoped \YBCO \cite{mook}, where neutron
scattering found period eight CDW but no static
spin order.
For slightly overdoped $\rm{Bi_2Sr_2CaCu_2O_{8+\delta}}$,
on which most of the tunneling 
experiments have been performed, neutron scattering experiments
suggest dynamic spin fluctuations \cite{fong}.
In our analysis we then assume that there is no
SDW order and concentrate on the effects of 
spin singlet translational symmetry breaking.
Another possible origin of a generalized
CDW with no spin symmetry breaking comes
from pinning of SDW by disorder \cite{polkovnikov2,SS}
or vortices \cite{yzhang,hujp}.

We restrict our analysis to the case of weak translational
symmetry breaking, 
when the new order parameter can be
treated as a small perturbation to the 
superconducting mean-field Hamiltonian. This limit clearly 
applies to the experimental situation
in \cite{davis1,kapitulnik1,davis2,davis3,kapitulnik2}, where the 
measured modulation is weak, and allows us to obtain explicit approximate
expressions for $\rho_{\bf Q}(\epsilon)$.
[This circumvents solving a complicated set of equations numerically,
as for instance carried out in~\cite{SS}.]
Furthermore, we do not address the issue of the origin of
charge order, but introduce it phenomenologically
and study its consequences for the STM experiments. Our basic motivation
is that a comparison of the energy dependence of $\rho_{\bf Q}$ with 
experimental data
can, in principle, be used to identify the correct order parameter(s) which,
in turn, is crucial for understanding their origin.  

This paper is organized as follows.  In Section II we introduce
mean-field Hamiltonians that describe several kinds of translational
symmetry breaking in a lattice system.  For these generalized CDWs
we derive an explicit formula for the Fourier component of the
tunneling density of states at the ordering wavevector. Section III
displays numerical results of this expression in the case of
\BSCCO type band structure and period four charge order. We show
that recent STM experiments by
\cite{kapitulnik1,kapitulnik2} are consistent with a generalized CDW that has
modulation of either the electron hopping or the pairing amplitude.
We also consider period eight structure that may be relevant to
$\rm{YBa_2Cu_3O_{6.35}}$. In Section IV we extend our analysis to phases with
randomness and show that a realistic value of the CDW correlation
length ($20 a_0$, with $a_0$ the unit cell size) provides good
agreement with the ``weak dispersion'' of the CDW peak observed in
\cite{kapitulnik2}.  As a different application of our formalism we
also consider localized perturbations in the crystal, such as impurity
potentials, and demonstrate that these can account for the strongly
dispersing peaks observed
in \cite{davis2,davis3} at wavevectors not corresponding to
the CDW order.
In Section V we review how to include a more realistic model
of the atomic wavefunctions, whose main effect
is to introduce a momentum dependent
structure factor. An important
implication of this result is that the signal at wavevectors differing
only by a reciprocal lattice vector should have peaks
at the same energies, although their amplitudes may differ.
We also discuss complications in the analysis
of the STM data introduced
by the normalization procedure used in the experiments.
Finally, in Section VI we discuss several sum rules for the Fourier components
of the density of states that may be useful for analyzing experiments.

\section{Order parameters and Mean-Field Hamiltonians
for generalized charge density wave phases}

\label{orderparameters}

Our starting point is a two dimensional one-band 
mean-field Hamiltonian that is commonly believed to be a good model
for the physics of the $d_{x^2-y^2}$ superconducting
state of the cuprates: 
\begin{eqnarray}
{\cal H} = \sum_{{\bf k}\sigma} \epsilon_{\bf k} c^\dagger_{{\bf
k}\sigma} c_{{\bf k}\sigma} + \sum_{{\bf k}} \Delta_{\bf k} (
c^\dagger_{{\bf k}\uparrow} c^\dagger_{-{\bf k}\downarrow} + c_{-{\bf
k}\downarrow} c_{{\bf k}\uparrow} )
\label{meand-field1}
\end{eqnarray}
Here 
$
\epsilon_{\bf k} = -2 t (\cos (k_x) + \cos (k_y))
- 4 t_1 \cos(k_x) \cos(k_y) -\mu
$,
$
\Delta_{\bf k} = \frac{\Delta_0}{2} ( \cos k_x - \cos k_y )
$
(from now on the unit cell size is set to unity),
$c_{{\bf r}\sigma}=N^{-1/2}\sum_{\bf k}
c_{{\bf k}\sigma} e^{ i {\bf k}{\bf r}}$,
and $N$ is the number of sites in the lattice.
The
Hamiltonian (\ref{meand-field1}) can be diagonalized using the 
Bogoliubov transformation 
$
c_{k\uparrow} = u_k \gamma_{k\uparrow} + v_k \gamma^\dagger_{-k\downarrow}
$,
$
c_{-k\downarrow} = u_k \gamma_{-k\downarrow} -
v_k \gamma^\dagger_{k\uparrow}
$
with $u_k^2+v_k^2=1$, $u_k v_k = { \Delta_k}/{2 E_k}$,
and $E_k = (\epsilon^2_k + \Delta^2_k)^{1/2}$.

Weak charge order may be introduced into the state (\ref{meand-field1})
by assuming the appearance of one or more of the
additional order parameters listed below. We note that distinction
between site and bond centered orders discussed below
is only defined for CDW  with integer periods.
\newline
{\it Site-centered charge density wave}. 
The local Hartree-Fock potential is
modulated along $x$ with potential extrema on the lattice sites
(see Fig. 1a):
\newline
$
\Delta {\cal H}_{1} 
= V_0 \sum_{\sigma, \, {\bf k}} 
\left(
c^\dagger_{{\bf k}+{\bf Q}\sigma} c_{{\bf k}\sigma}
+ c^\dagger_{{\bf k}\sigma} c_{{\bf k}+{\bf Q}\sigma} \right)
$
\newline
{\it Bond-centered charge density wave.} 
The local Hartree-Fock potential is
modulated along $x$ with the extrema of the modulation
at midpoints of the horizontal bonds (see Fig. 1b):
\newline
$
\Delta {\cal H}_{2} 
= V_0 \sum_{\sigma, \, {\bf k}} \left( \alpha^* \,
c^\dagger_{{\bf k}+{\bf Q}\sigma} c_{{\bf k}\sigma}
+ \alpha\, c^\dagger_{{\bf k}\sigma} c_{{\bf k}+{\bf Q}\sigma} \right)
$, where $\alpha = e^{ i Q/2}$.
\newline
{\it Longitudinal dimerization.} 
Single electron tunneling amplitudes are modulated
on the horizontal bonds and the wavevector of modulation
is along the same direction ({\em i.e.} along $x$).
The bond centered version, in which the extrema of the
modulation lie on the bonds (see Fig. 1c) corresponds to:
$
\Delta {\cal H}_{3} 
=  V_0 \sum_{\sigma, \, {\bf k}}
\cos (k_x+\frac{Q}{2}) \left( \alpha^* \,
c^\dagger_{{\bf k}+{\bf Q}\sigma} c_{{\bf k}\sigma}
+ \alpha\, c^\dagger_{{\bf k}\sigma} 
c_{{\bf k}+{\bf Q}\sigma} \right)
$.
\newline
{\it Transverse dimerization}.
Single electron hopping is modulated
on the vertical bonds, and the wavevector of modulation
is along the horizontal direction ({\em i.e.} along $x$).
The site centered version ({\em i.e.} with extrema of the
modulation realized on the vertical bonds) is shown  in Fig. 1d,
and corresponds to:
\newline
$
\Delta {\cal H}_{4} 
=  V_0 \sum_{{\bf k}\sigma} \cos k_y 
\left(
c^\dagger_{{\bf k}+{\bf Q}\sigma} c_{{\bf k}\sigma}
+ c^\dagger_{{\bf k}\sigma} c_{{\bf k}+{\bf Q}\sigma} \right)
$.
\newline
{\it Anomalous longitudinal} {\it dimerization}. 
The $x$-components of the $d_{x^2-y^2}$-wave pairing amplitudes are
modulated in the $x$ direction.
The bond centered version, shown in Fig. 1c, corresponds to:
$
\Delta {\cal H}_{5} 
=  
V_0 \sum_{\sigma, \, {\bf k}} \large\{
\cos (k_x+\frac{Q}{2}) \left(\, \alpha^*\,
c^\dagger_{{\bf k}+{\bf Q}\uparrow} 
c^\dagger_{-{\bf k}\downarrow}
+ \, \alpha\,
c^\dagger_{{\bf k}\uparrow} c^\dagger_{-{\bf k}-
{\bf Q}\downarrow} \right)
+ h.c. \large\}
$.
\newline
{\it Anomalous transverse dimerization}. 
The $y$-components of the
$d_{x^2-y^2}$-wave pairing amplitudes are
modulated in the $x$ direction.
The site centered version, shown in Fig. 1d, corresponds to:
$
\Delta {\cal H}_{6} 
= V_0 \sum_{\sigma, \, {\bf k}} \large\{
\cos k_y \left(\,
c^\dagger_{{\bf k}+{\bf Q}\uparrow} 
c^\dagger_{-{\bf k}\downarrow}
+ 
c^\dagger_{{\bf k}\uparrow} c^\dagger_{-{\bf k}-
{\bf Q}\downarrow} \right)
+ h.c. \large\}
$.

Note that these subdominant order parameters may appear either as a
result of a phase transition in the bulk, or due to pinning by
vortices, impurities or any other defects (see discussion in Section
IV).  Following experimental observations in
\cite{kapitulnik1,kapitulnik2,mook00}, we assume that the order is
unidirectional, and choose the ordering wavevector to be ${\bf Q}= Q
\hat{e}_x$ \cite{note}.  However,
even if we were to assume checkerboard order, our analysis is carried
out to linear order in perturbation theory and, by linear
superposition, our results would be identical to those obtained
for unidirectional order.
For the STM experiments in
\BSCCO
\cite{davis1,kapitulnik1},
$Q=2\pi/4$, while the neutron scattering experiments on \YBCO
\cite{mook} correspond to the smaller value: $Q=2\pi/8$. 
We point out that  the six cases listed above
are, in general, not orthogonal to each other in a symmetry sense.
As a result it is conceivable that more than one order parameter
could be simultaneously non-zero; for
example, in a microscopic model without particle-hole
symmetry, a simple CDW would be expected
to induce dimerization as the two order parameters are linearly coupled
\cite{kivelson2}.


Upon expressing
the Hamiltonians $\Delta {\cal H}$ in the basis of
Bogoliubov quasiparticles, they reduce to the generic form
\begin{eqnarray}
\Delta {\cal H}_i &=& \sum_{k\sigma} 
\left[
A^{i}_k \gamma^\dagger_{{\bf k}\sigma} \gamma_{{\bf k}+{\bf Q}\sigma}
+A^{i*}_{\bf k} \gamma^\dagger_{{\bf k}+{\bf Q}\sigma} \gamma_{{\bf k}\sigma}
\right]
\nonumber\\
&+& \sum_k
\left[
B^{i}_k \gamma^\dagger_{{\bf k}\uparrow} \gamma^\dagger_{{\bf k}+{\bf Q}\downarrow}
+B^{i*}_{\bf k} \gamma^\dagger_{{\bf k}+{\bf Q}\uparrow} \gamma^\dagger_{{\bf k}\downarrow}+{\rm h.c.}
\right],
\label{genpert}
\end{eqnarray}
where
\begin{eqnarray}
\begin{array}{c l}
A_{\bf k}^1 = V_0\omega_{\bf k}
&B_{\bf k}^1 = V_0\eta_{\bf k}
\\
A_{\bf k}^2 = V_0\alpha \,\omega_{\bf k}
&B_{\bf k}^2 = V_0\alpha\,\eta_{\bf k}
\\
A_{\bf k}^3 = V_0\alpha \,\cos(k_x+\frac{Q}{2})\,\omega_{\bf k} 
&B_{\bf k}^3 = V_0\alpha\,\cos(k_x+\frac{Q}{2})\,\eta_{\bf k}
\\
A_{\bf k}^4 = V_0\cos(k_y)\,\omega_{\bf k}
&B_{\bf k}^4 = V_0\cos(k_y)\,\eta_{\bf k}
\\
A_{\bf k}^5 = -V_0\alpha \,\cos(k_x+\frac{Q}{2})\,\eta_{\bf k}
&B_{\bf k}^5 = V_0\alpha\,\cos(k_x+\frac{Q}{2})\,\omega_{\bf k}
\\
A_{\bf k}^6 = -V_0\cos(k_y)\,\eta_{\bf k}
&B_{\bf k}^6 = V_0\cos(k_y)\,\omega_{\bf k}
\end{array}
\label{Vkexpr}
\end{eqnarray}
in terms of the coherence factors
$\omega_{\bf k}=u_{{\bf k}+{\bf Q}} u_{\bf k} - 
v_{{\bf k}+{\bf Q}} v_{\bf k}$ and $\eta_{\bf k}=
u_{{\bf k}+{\bf Q}} v_{\bf k}+ v_{{\bf k}+{\bf Q}}u_{\bf k}$.

STM experiments measure the local density of states
$
\rho({\bf r},\epsilon)=\sum_{n\sigma} 
\left\{ 
|\langle n |c^\dagger_{{\bf r}\sigma} | 0 \rangle |^2
\delta (\epsilon - \epsilon_{n0}) 
+\right.$ $\left.
|\langle n |c_{{\bf r}\sigma} | 0 \rangle |^2
\delta (\epsilon + \epsilon_{n0})
\right\}
$,
where the summation over $n$ ranges over all excited states.
In particular, we are interested in the Fourier transform
\begin{eqnarray}
\rho_{\bf q}(\epsilon) &=& \frac{1}{N} \sum_{\bf r} e^{-i {\bf q}{\bf r}} \rho({\bf r},\epsilon)
\nonumber\\
&=&\frac{1}{N}\sum_{n{\bf k}\sigma}\left[ 
\langle 0 | c_{{\bf k}+{\bf q}\sigma} | n \rangle
\,\langle n |  c^\dagger_{{\bf k}\sigma} | 0 \rangle
\delta (\epsilon - \epsilon_{n0})
\right.\\
&\,&\qquad\left.
\,\,+\,\langle 0 | c^\dagger_{{\bf k}\sigma} | n \rangle
\,\langle n |  c_{{\bf k}+{\bf q}\sigma} | 0 \rangle
\delta (\epsilon + \epsilon_{n0})
\right].\nonumber
\end{eqnarray}
Although a full treatment of all terms in (\ref{genpert}) is
complicated, progress can be made if we assume that the ordering 
represented by $\Delta {\cal H}$ is weak, allowing
us to obtain an analytic expression for the Fourier transform
that is exact to linear order in $V_0$.
This is then the sum of two contributions
\begin{eqnarray}
\rho_{\bf Q}(\epsilon)=\rho_{\bf Q}^{\rm A}(\epsilon)+\rho_{\bf Q}^{\rm B}(\epsilon)+O(V_0^2),
\label{lincomb}
\end{eqnarray}
where $\rho_{\bf Q}^{\rm A}(\epsilon)$
  is obtained by ignoring the
$B_{\bf k}$  term in the perturbation (\ref{genpert})
and {\it vice versa}.  

A small value of $V_0$ leads to another important simplification:
We only need to consider the pairwise mixing between states
connected by  $\Delta {\cal H}$.
For instance, in computing $\rho_{\bf Q}^{\rm A}(\epsilon)$
for $Q=2\pi/4$
one would have to analyze coupled equations for
four quasiparticles   
(${\bf k}$, ${\bf k}+{\bf Q}$, ${\bf k}+2 {\bf Q}$,
and ${\bf k}+ 3 {\bf Q}={\bf k}-{\bf Q}$) connected by the perturbation. 
However, in the limit when $V_0$ is small, there
is at most one pair of quasiparticles 
that have similar energies, and that will
be hybridized appreciably by $\Delta {\cal H}$. This hybridization
can be analyzed by 
diagonalizing the corresponding two-by-two Hamiltonian, which
gives the new eigenstates $| \alpha_{\bf k \sigma} \rangle$
and $ | \beta_{\bf k \sigma} \rangle$ with energies
$
\tilde{E}_{{\bf k}\pm} =
\frac{E_{\bf k}+E_{{\bf k}+{\bf Q}}}{2} \pm
\left[(\frac{E_{\bf k}-E_{{\bf k}+{\bf Q}}}{2})^2 + 
|A_{\bf k}|^2\right]^{1/2}
$. Note that these states satisfy
$
\langle 0 | c_{{\bf k}+{\bf Q}\sigma} | \alpha_{\bf k \sigma} \rangle
\langle \alpha_{\bf k \sigma} | c^\dagger_{{\bf k}\sigma} | 0 \rangle
= \frac{1}{2}\,u_{\bf k} u_{{\bf k}+{\bf Q}} \sin 2\theta_{\bf k} e^{-i \chi_{\bf k}}
$,
$
\langle 0 | c_{{\bf k}+{\bf Q}\sigma} | \beta_{\bf k \sigma} \rangle
\langle \beta_{\bf k \sigma} | c^\dagger_{{\bf k}\sigma} | 0 \rangle
= - \frac{1}{2}\, 
u_{\bf k} u_{{\bf k}+{\bf Q}} \sin 2\theta_{\bf k} e^{-i \chi_{\bf k}} 
$, where we have defined $A_{\bf k} = |A_{\bf k}|\,e^{i \chi_{\bf k}}$,
and 
$ \tan 2 \theta_{\bf k} = 2 |A_{\bf k}|/
(E_{\bf k}-E_{{\bf k}+{\bf Q}})$.
From these results one easily finds
\begin{eqnarray}
\rho_{\bf Q}^{\rm A}(\epsilon) &=& \frac{1}{N} \sum_{{\bf k}}
\frac{A^*_{\bf k}}
{\sqrt{\left(\frac{E_{\bf k}-E_{{\bf k}+{\bf Q}}}{2}\right)^2 + 
|A_{\bf k}|^2}} 
\nonumber\\ 
 &\times&\left[
u_{\bf k} u_{{\bf k}+{\bf Q}} \left(
\delta (\epsilon - \tilde{E}_{{\bf k}+})-
\delta (\epsilon - \tilde{E}_{{\bf k}-}) \right)
\right.
\label{Afinal}\\
 &\,&\left.
+v_{\bf k} v_{{\bf k}+{\bf Q}} \left(
\delta (\epsilon + \tilde{E}_{{\bf k}+})-
\delta (\epsilon + \tilde{E}_{{\bf k}-}) \right)
\right]
\nonumber
\end{eqnarray}

When considering $\rho_{\bf Q}^{\rm B}$, one would naively expect that it is
always smaller than $\rho_{\bf Q}^{\rm A}$, because the perturbation terms of
the form $\gamma^\dagger_{{\bf k} \pm{\bf Q}\uparrow}
\gamma^\dagger_{-{\bf k}\downarrow}$ connect states that differ
in energy by $E_{\bf k}+E_{{\bf k}+{\bf Q}}$, a factor that is never
small. However, in some cases the coherence factors in $A_{\bf k}$
vanish at important regions of the Brillouin zone, making $\rho_{\bf
Q}^{\rm A}(\epsilon)$ anomalously small.  In addition, as we discuss below,
both $\rho^{{\rm A},{\rm B}}(\epsilon)$ are large at biases corresponding to the
saddle points on the degeneracy lines $E_{\bf k} = E_{{\bf k}+{\bf
Q}}$ and van Hove singularities of the Bogoliubov quasiparticles
$\epsilon \approx \Delta_0$. A nearly identical analysis
of the one above for $\rho^{\rm A}$ yields
\begin{eqnarray}
\rho_{\bf Q}^{\rm B}(\epsilon) &=& \frac{1}{N} \sum_{{\bf k}}
\frac{B^*_{\bf k}}
{\sqrt{\left(\frac{E_{\bf k}+E_{{\bf k}+{\bf Q}}}{2}\right)^2 + 
|B_{\bf k}|^2}} 
\nonumber\\ 
 &\times
 &\left[
 u_{\bf k} v_{{\bf k}+{\bf Q}}
\delta (\epsilon - \hat{E}_{{\bf k}+})
+u_{{\bf k}+{\bf Q}} v_{\bf k}
\delta (\epsilon - \hat{E}_{{\bf k}-}) \right.
\label{Bfinal}\\
 &\,&\left.
-u_{{\bf k}+{\bf Q}} v_{\bf k}
\delta (\epsilon + \hat{E}_{{\bf k}+})
-u_{\bf k} v_{{\bf k}+{\bf Q}}
\delta (\epsilon + \hat{E}_{{\bf k}-}) \right]
\nonumber
\end{eqnarray}
where $\hat{E}_{{\bf k}\pm}=\pm\frac{E_{\bf k}-E_{{\bf k}+{\bf Q}}}{2} +
\left[(\frac{E_{\bf k}+E_{{\bf k}+{\bf Q}}}{2})^2 + 
|B_{\bf k}|^2\right]^{1/2}$.
Equations~(\ref{Afinal},\ref{Bfinal}), are two key 
results of this paper.
In combination with Eqns (\ref{Vkexpr}),
they provide an explicit expression for the 
energy dependence of the Fourier
component of the local density of states 
$\rho_{\bf Q}(\epsilon)$ when the translational symmetry breaking is 
weak.

From the form of $A_k$ and $B_k$, it is obvious that when there is no
mixing between bond and site centered CDW, $\rho_{\bf Q}(\epsilon)$
can be made real at all energies by an appropriate choice of the
overall phase, {\it i.e.} by a shift in the origin
of coordinates when doing the Fourier
transform.  One obvious observation is that the results for the
site-centered and bond-centered CDW are identical modulo an overall
phase factor of $e^{i\,Q/2}$.  If one defines the Fourier transform in
such a way that it is real in both cases, the origin will coincide
with one of the sites of the lattice for the site-centered CDW, and it
will be at the center of a bond for the bond-centered CDW.  Hence
careful analysis of the STM data allows one to distinguish two kinds
of CDW, a task that is not possible in neutron scattering experiments
with current resolution. Mixing site and bond-centered
orders breaks inversion symmetry and leads to a
complex-valued $\rho_{\bf Q}(\epsilon)$.

\section{Charge order with no randomness}

\subsection{Period four CDW in \BSCCO}

We first focus on modulations at ${\bf Q}=(2\pi/4,0)$
that is relevant to \BSCCO \cite{davis1,kapitulnik1}.
Figure \ref{fig2} shows results of the numerical
evaluation of formulas
(\ref{lincomb}), (\ref{Afinal}) and (\ref{Bfinal}) for various perturbations
(\ref{Vkexpr}).
[As transverse and longitudinal dimerization curves are qualitatively
similar, curves corresponding to the former are not displayed.]
We choose the band structure and the value of $\Delta_0$ 
in (\ref{meand-field1}) appropriate to
\BSCCO:
$t_1/t=-0.3$, $\mu/t=-0.99$ (this corresponds to $14\%$ doping),
$\Delta_0/t=0.14$ and $\Delta_0=40$ meV \cite{shen}. We set
$V_0/t=0.02$, although its precise value is inconsequential, as
$\rho_{\bf Q}(\epsilon)$ scales linearly with $V_0$ when the latter is
sufficiently small.

If we turn our attention to the expression for $\rho^{\rm A}$,
Eqn.~(\ref{Afinal}), we see that the energy denominator is smallest
for those quasiparticles lying close to the degeneracy points $E_{\bf
k}=E_{{\bf k}+{\bf Q}}$, which are strongly hybridized by the $A_k$
part of the perturbation.  Figure~(\ref{fig4}) shows the four loci of
such points, $a$ through $d$, that are degenerate with $a'$ to $d'$
respectively.  The pairs $aa'$ and $bb'$ are obvious, since they have
$k_x= \pm \pi/4$ and $k_x=\pm 3 \pi/4$ (for the same $k_y$); the other
two require a more detailed analysis of the band structure.  Out of
the set of degeneracy points, we expect large contributions from the
neighborhood of points A and B, as the dispersion of hybridized
energies $\tilde{E}_{{\bf k}\pm}$ is flat at these points.  These same
regions of the Brillouin zone will dominate the $\rho_B$ contribution,
since the energy denominator in
(\ref{Bfinal}) will be small only if both ${\bf k}$ and ${\bf k}+{\bf
Q}$ lie close to the Fermi surface, which occurs only in the
neighborhood of points A and B.  In addition we expect, for both
$\rho^{\rm A}$ and $\rho^{\rm B}$ pieces, a large contribution at
$\epsilon=\Delta_0$, where a van Hove singularity 
for the Bogoliubov quasiparticles yields a logarithmic
divergence in the density states.

We turn now to the numerical results displayed in Fig~\ref{fig2}.
Consider first the simple CDW curves.
The sharp features that dominate the
CDW plots can be understood in terms of the degeneracies mentioned
above:
the peak at energies around
$0.5 \Delta_0$ comes from the vicinity of the A point,
the peak around $0.7 \Delta_0$ comes from the vicinity of
B, and the pile around $\Delta_0$ comes from
the van Hove singularity near the $(0,\pi)$ and $(\pi,0)$ points.
The longitudinal dimerization results can be similarly understood
by taking into account the additional minus sign
in the vicinity of the point B due to the
$\cos(k_x+\pi/4)$ factor in $A_{\bf k}$ and $B_{\bf k}$.
The results for the anomalous dimerization can also be understood in this
framework after taking into account the extra sign 
modulation in
$u_{\bf k} v_{\bf k}$, which changes sign whenever
$\Delta_{\bf k}$ does.
Note that, for all perturbations considered,
$\rho_{\bf Q}(\epsilon)$ displays approximate particle-hole
symmetry for small biases, as observed in STM measurements
\cite{kapitulnik1}. This is not a generic property
of $\rho_{\bf Q}(\epsilon)$; for example, for a diamond-shaped Fermi
surface $\mu=t'=0$ the CDW signal is exactly antisymmetric. Finally, note
that $\rho_{\bf Q}(\epsilon)$ goes to zero at $\epsilon=0$ in all
cases; this reflects the vanishing density of low-energy quasiparticle
states in an ideal $d$-wave superconductor.

While the results in Figure~\ref{fig2} describe a system with
infinite quasiparticle lifetime and no disorder,
in a real system disorder will smear the sharp features in 
$\rho_{\bf Q}(\epsilon)$.  To model this, these curves are
re-displayed in Figure~\ref{fig3}
after smearing over an energy width
$w=0.2 \Delta_0$.
This procedure smooths the sharp features in the spectra, and
generates finite intensity at low energies.  Notice that the smeared
CDW curve does not have the two large peaks surrounding zero bias, nor
does it have clear zero crossings at $|\epsilon|\approx\Delta_0$, the
dominant features of the STM spectra observed in
\cite{davis1,kapitulnik1}.  By contrast, the signal related to
longitudinal dimerization or, especially, to either kind of anomalous
dimerization, share many of the qualitative properties of the data.
However, neither curve by itself accounts for all the observed
features in the data.  This prompts us to consider a combination of
several kinds of order.  For example, if we assume that no pairing
modulation is present, the combination of longitudinal dimerization
and CDW, $({\rm long.dim.})+1.05({\rm CDW})$, shown as a solid curve
in Figure~\ref{fig5} reproduces the STM results reasonably well, with
only a small difference in the position of the peaks ($\pm 17$ meV,
compared to experimentally observed $\pm 25$ meV).  Alternatively we
can match experimental data by considering the combination of
anomalous longitudinal dimerization and CDW, $({\rm
anom.long.dim})+0.2({\rm CDW})$, shown as a dashed curve in
Figure~\ref{fig5}. It slightly overestimates the peak bias to be $\pm
29$ meV, and yields a low intensity at zero bias. Any intermediate
combination between these two scenarios also gives good agreement with
experiments.  Although CDW was used in both combinations discussed
above, it can be substituted by transverse dimerization, which yields
a qualitatively similar $\rho_{\bf Q}(\epsilon)$ to CDW.  We note
that, for $\epsilon
\leq 3\Delta_0$, the results come from the vicinity of the Fermi
surface and are robust against variations in the band structure that
do not alter qualitatively the shape of the Fermi surface (e.g. the
$a$ and $b'$ lines do not move below the Fermi surface).

We note, however, that a certain care should
be exercised when comparing our
results to the STM spectra
in \cite{davis1,kapitulnik1,davis2,davis3,kapitulnik2}.
An additional complication of the experiments
is that for every point on the surface of the sample
the height of the STM tip is adjusted
to keep the tunneling current at a certain
voltage fixed. This implies that
the local density of states is not measured directly,
but instead its product
with some space dependent function is measured. In Section
\ref{Experiments} we review how this normalization procedure 
can be included in analysis.

\subsection{Period eight CDW in \YBCO}

To model \YBCO for which CDW-type peaks
have been observed at $Q=2\pi/8$ \cite{mook}
we take the same band structure $t_1/t=-0.3$, but a different value
of the chemical potential $\mu/t=-0.815$ (this
corresponds to $6\%$ doping). We use
the same value of $\Delta_0/t=0.14$, $\Delta_0=40$ meV,
$V_0/t=0.02$, and keep the energy smearing $w=0.2 \Delta_0$.
The main difference with the charge order at $Q=2\pi/4$ is that the analog of
line {\it a} in this case is inside the Fermi surface, so that the only
contributions will come from the vicinity of point B at energies around 
$0.8\Delta_0$.  
This leads to less structure 
in $\rho_{\bf Q}(\epsilon)$ and smaller intensity at zero energy
(see Fig. \ref{fig6}).

\section{Dispersion of the STM spectra}

Recent experiments \cite{davis2,davis3} demonstrated
that the STM spectra  of \BSCCO cannot be explained
by charge order at a unique wavevector.
Peaks in $\rho_{\bf q}(\epsilon)$ have been observed
away from $(2\pi/4,0)$ and the
wavevectors of the peaks are energy dependent. In this section we
review and compare two possible scenarios for such 
dispersion of the STM spectrum: 1) randomness and pinning of 
charge order, 2) scattering of BCS quasiparticles by impurities
and crystal defects. Both cases can be described using an extension of the
formalism presented in the previous section.  We
consider a single particle Hamiltonian that generalizes equation 
(\ref{genpert})
\begin{eqnarray}
\Delta {\cal H}&=&\sum_{{\bf kq}\sigma}\left[
V_{{\bf kq}}\, c_{{\bf k},\sigma}^\dagger c_{{\bf k}+{\bf q},\sigma}^{\,}+{\rm h.c.}
\right.\label{general}
\\&\,&\left.
+W_{{\bf kq}} \left( c_{{\bf k},\sigma}^\dagger c_{-{\bf k}-{\bf q},-\sigma}^\dagger+
c_{-{\bf k},-\sigma} c_{{\bf k}+{\bf q},\sigma}
\right) + {\rm h.c.}\right].
\nonumber
\end{eqnarray}
Here ${\bf q}$ describes the wavevector of the potential modulation,
and the ${\bf k}$ dependence of $V$ and $W$ gives its
internal structure ({\it e.g.} simple CDW vs dimerization)
\cite{singlet}.
In Section II we considered charge order at a unique
wavevector that corresponds to taking  
potentials $V$ and $W$
as $\delta({\bf q}-{\bf Q})$. In the case of a
disordered CDW we expect that these functions are no longer
$\delta$-functions but are centered
narrowly around some particular wavevector. By contrast,
when translational symmetry breaking comes from impurities,
we expect to find $V$ and $W$ that extend
over a wide range of wavevectors ${\bf q}$.
A crucial property of (\ref{Afinal}) and (\ref{Bfinal})
is that $\rho_{\bf q}(\epsilon)$ depends linearly
on the perturbations $V_{\bf q}$ and $W_{\bf q}$, 
hence the formalism for computing $\rho_{\bf q}(\epsilon)$
can be applied independently to each wavevector ${\bf q}$.


The charge order observed in \BSCCO \cite{kapitulnik1,kapitulnik2}
had strong signatures of randomness
and pinning in the form of lattice defects. The correlation
length estimated from the distance between defects was 
$\approx 20 a_0$. If we assume the charge order
to be of the form 1.05(CDW)+(long.dim.), we can describe it
as
\begin{eqnarray}
V_{\bf kq}=V_0({\bf q})\left(1.05+\cos(k_x+\frac{q}{2})\right),\quad W_{\bf kq}=0
\label{specific}
\end{eqnarray}
where $V_0({\bf q})$ is a Gaussian distribution function centered
at $(2\pi/4,0)$ with a width $2\pi/20 a_0$. 
We display in Fig.~\ref{fig7} the signal produced by a perturbation
of this kind, for bias voltages 8, 12, 16, and 20 mV,
as a function of wavevectors along the $(0,0)$ to $(\pi,0)$
direction.
The resulting dispersion agrees closely with that observed in
\cite{davis2} and \cite{kapitulnik2}.


In experiments of Hoffman {\it et.al} \cite{davis2}
and McElroy {\it et.al} \cite{davis3} peaks
in the  LDOS were observed at very different wavevectors
from $(2\pi/4,0)$ (including some in diagonal directions).
This suggest that  either $V_{\bf kq}$ or $W_{\bf kq}$ must be non-zero
over a fairly wide range of values of {\bf q},
and the most natural candidate is scattering by impurities 
\cite{davis2,davis3,dhlee}.
For concreteness, we assume that the impurity induces a higher
chemical potential at a single site, so the perturbation
used corresponds to a simple CDW which is uniform in {\bf q},
$V_{\bf kq}=V_0,$ $W_{\bf kq}=0$.
In Fig.~\ref{fig8} we show the signal computed along the $(0,0)$
to $(\pi,0)$ direction at bias voltages 8, 12, 16, and 20 mV.
In all cases there is a pronounced peak
that disperses with the applied bias voltage.
To find the positions of these peaks we reverse
the arguments given in Section II. There, we started with
a potential at wavevector ${\bf Q}$ and found that
only quasiparticles at certain energies were
strongly affected by it. Now we need to find 
the modulation wavevector that affects quasiparticles at
a given energy. From the band structure of
\BSCCO we find for the peak positions 
(in units of $2\pi$):
0.35, 0.32, 0.29, and 0.26. The curves on
Fig.~\ref{fig8} show general agreement with this
``quasiparticle scattering'' argument \cite{davis2,davis3},
except for a consistent small shift to lower
wavevector, which comes from the energy smearing procedure.
This dispersion is stronger than that displayed in the data
at wavevector $(2\pi/4,0)$, but is in good agreement
with the dispersion observed at other wavevectors.


In the discussion above we considered two situations: ordered CDW and
non-interacting electrons with impurities.  There may also be an
intermediate regime with interacting electrons close to the CDW
instability and with disorder \cite{polkovnikov2,kivelson}.
Qualitatively, this case may be described by equation
(\ref{general}) but with the potentials $V_{\bf kq}$ and $W_{\bf kq}$
coming not only from the external fields but also from the density
induced in the electron system. For simplicity let's take
only one of the channels discussed in Section \ref{orderparameters},
e.g. simple CDW (the generalization to the case of several
channels is straightforward). Then
\begin{eqnarray}
\Delta {\cal H} = 
\sum_{\bf q}\,
\hat{\rho}^\dagger_{\bf q}U_{\bf q}\hat{\rho}_{\bf q}+
\sum_{\bf q}\,
V^{\rm ext}_{\bf q}\hat{\rho}^\dagger_{\bf q}+{\rm h.c.}
\label{interimp}
\end{eqnarray}
with 
\begin{eqnarray}
\hat{\rho}_{\bf q}= \sum_{{\bf k}\sigma} 
c_{{\bf k}\sigma}^\dagger c_{{\bf k}+{\bf q}\sigma}
\nonumber
\end{eqnarray}
Response of the quasiparticles to (\ref{interimp})
in the Hartree approximation is determined by the
effective perturbation Hamiltonian
\begin{eqnarray}
\Delta {\cal H} &=& 
\sum_{{\bf kq}\sigma}\, V^{\rm eff}_{\bf q} 
c_{{\bf k}+{\bf q}\sigma}^\dagger c_{{\bf k}\sigma}
+{\rm h.c.}
\nonumber\\
V^{\rm eff}_{\bf q} &=& 
 V^{\rm ext}_{\bf q} + U_{\bf q} 
\langle \hat{\rho}_{\bf q} \rangle
=
\frac{V^{\rm ext}_{\bf q}}{1-U_{\bf q}\chi_0({\bf q},\omega=0)}
\label{interimp3}
\end{eqnarray}
where $\chi_0({\bf q},\omega=0) 
= \langle \hat{\rho}_{\bf q} \hat{\rho}^\dagger_{\bf q}\rangle$ 
must be computed for the non-interacting system.  This corresponds to
contributions to the LDOS from the class of diagrams
in figure \ref{fig9}(a).
When the system is close to the CDW instability,
the denominator of (\ref{interimp3}) approaches zero around some
particular wavevector ${\bf Q}$.  Hence, $V^{\rm eff}_{\bf q}$ may be
peaked around ${\bf Q}$ even when $V^{\rm ext}_{\bf q}$ is momentum
independent.  In principle we can go beyond the Hartree approximation,
including diagrams such as those shown in figure \ref{fig9}(b,c).
These will introduce a frequency-dependent self energy, as in the case of pinned spin
density fluctuations considered in \cite{SS}.

It is interesting to ask whether by analyzing experimental data one
can separate contributions of disordered CDWs from those due
to quasiparticle
scattering off impurities. Reference \cite{kapitulnik2} pointed
out that the presence of a weakly dispersing signal at 
wavevector $(2\pi/4,0)$ makes charge order
a likely candidate at that wavevector.
Here we build on this idea and suggest that a more
consistent approach is to analyze
$\rho_{\bf q}(\epsilon)$ at many different wavevectors using a reasonable
set of basis functions, {\it e.g.} simple CDW and dimerization. Such
analysis will give the ${\bf q}$ dependence of various components 
in the potentials $V_{\bf kq}$ or $W_{\bf kq}$.
We expect that some of them will be almost uniform in momentum space
and  correspond to localized defects, such as impurities;
whereas others will be centered around particular
wavevectors and arise from the existence or at least proximity
of charge order.


When the field of view of the 
STM measurement contains more than one
impurity there are several important questions that we need to
address. We must ask how the contributions from different impurities
add up and whether the system retains the property of uniformity of
phase of $\rho_{\bf q}(\epsilon)$ at fixed ${\bf q}$ but different
$\epsilon$. We consider impurities that cause an arbitrary potential
$V_{\bf kq}$ in Eqn. (\ref{general}), {\it i.e.} they may modify the chemical
potential, the electron kinetic energy, or the pairing amplitude, but
first we assume that all impurities are identical. 
The Fourier component of the LDOS is proportional
to $ V^{tot}_{\bf kq}= V_{\bf kq} \sum_{{\bf r}_a} e^{i {\bf q
r}_a} $, where ${\bf r}_a$ runs over impurity positions.  If one
impurity does not break parity symmetry, we can make $V_{\bf kq}$ real
by choosing the origin at the position of this impurity (see also
discussion in Section II). This implies that $V^{tot}_{\bf kq}$ has a phase
that depends on ${\bf q}$ only and is the same for all ${\bf k}$,
which in turn proves
that, at fixed ${\bf q}$, $\rho_{\bf q}(\epsilon)$ has constant phase
(modulo $\pi$) for all values of $\epsilon$ (see Eqns (5)-(9)). 
So, in the case of
identical impurities we have only one phase to worry about and we can
always make $ \rho_{\bf q}(\epsilon)$ real by an appropriate choice of
origin.  When impurities are different, we will have an intrinsically
complex $ \rho_{\bf q}(\epsilon)$, with possibly energy dependent
phase at different bias voltages. In either case,
interference among the impurities
leads to an appreciable suppression of the amplitude of
$\rho_{\bf q}(\epsilon)$. When there are many impurities in the area
$A$ of the STM field of view, and their positions are uncorrelated, 
each impurity introduces a random phase to $V^{tot}_{\bf kq}$, whose amplitude
can be analyzed in terms of a random walk.  Therefore,
in a typical experiment we expect that with increasing the
system size, $\langle \left| \rho_{\bf q}(\epsilon) \right| \rangle$ 
will decay as $1/\sqrt{A}$,
with statistical
fluctuations of the same order. This argument also applies to the
case of disordered CDW, where the role of impurities is played by defects
in the CDW lattice.

\section{Experimental Considerations}
\label{Experiments}

Our discussion in the earlier sections was restricted to models on a
square lattice for which we calculated the lattice density of states
in the presence of several kinds of translational symmetry breaking.
In analyzing actual experimental data, additional effects need
to be taken into account: the real space structure of
the atomic wavefunctions, and the current normalization condition used in
the STM measurements 
\cite{davis1,kapitulnik1,davis2,davis3,kapitulnik2}. These are reviewed below.

\subsection{Structure factors}

For lattice Hamiltonians, wavevectors that differ by the reciprocal
lattice vectors ${\bf G}$ are equivalent. For the Fourier components
of the local density of states $\rho_{\bf q}$ this implies that $
\rho_{{\bf q}+{\bf G}}^{\rm lattice}(\epsilon)
= 
\rho_{{\bf q}}^{\rm lattice}(\epsilon)
$
for any ${\bf G}$. To understand why this equivalence is
not observed in experiments we must take into account the real space structure 
of the Wannier wavefunctions of electrons in the conduction band.  
Here we study the effects of a
single-band tight-binding model, in which Bloch states at the Fermi level
can be written as a superposition of localized atomic
orbitals 
$
\psi_{\bf k}({\bf r})=\sum_{\bf R}e^{i {\bf kR}}\phi({\bf r}-{\bf R})
$.

We begin by projecting the Bogoliubov-de Gennes (BdG) wavefunctions
in terms of the single-band
wavefunctions,
\begin{eqnarray*}
u_n({\bf r})&=&\sum_{\bf k} a_{\bf k}^n\psi_{\bf k}({\bf r}),\\
v_n({\bf r})&=&\sum_{\bf k} b_{\bf k}^n\psi^*_{\bf k}({\bf r}).\\
\end{eqnarray*}
If we know how an operator $\hat{\Theta}$ acts on the Bloch wavefunctions $\psi_{\bf k}$,
$\hat{\Theta} \psi_{\bf k}=\sum_{{\bf k}'} \Theta_{{\bf kk}'} \psi_{{\bf k}'}$, then
the above relation induces an action on $a_{\bf k}$
through $\hat{\Theta} a_{\bf k}=\sum_{{\bf k}'} a_{{\bf k}'} \Theta_{{\bf k}'{\bf k}}$
(and similarly for $b_{\bf k}$).  Thus, the solutions to the BdG equation
\begin{eqnarray*}
\left(
\begin{array}{c c}
\hat{\xi}+\hat{V} & \hat{\Delta}+\hat{W} \\
\hat{\Delta}^\dagger+\hat{W}^\dagger & -\hat{\xi}-\hat{V} 
\end{array}\right)
\left(\begin{array}{c} a_{\bf k}^n \\ b_{\bf k}^n \end{array}\right)
=E_n \left(\begin{array}{c} a_{\bf k}^n \\ b_{\bf k}^n \end{array}\right)
\end{eqnarray*}
will be independent of the Wannier wavefunction $\phi({\bf r})$
once we determine the action of the BdG
operator on the Bloch wavefunctions $\psi_{\bf k}$.

For positive biases $\epsilon>0$ (the $\epsilon<0$ case can be analyzed
analogously), the LDOS is given by 
\begin{eqnarray*}
\rho^{\rm phys}_{\bf q}(\epsilon)&=&\int d^2{\bf r} \,e^{i{\bf qr}} \sum_n u_n^*({\bf r})u_n({\bf r}) \delta(\epsilon-E_n)\\
 &=&\sum_n\delta(\epsilon-E_n) \sum_{{\bf kk}'}a_{\bf k}^{*n}a_{{\bf k}'}^{n}J({\bf k},{\bf k}',{\bf q}),
\end{eqnarray*}
where
\begin{eqnarray*}
J({\bf k},{\bf k}',{\bf q})&=&\int d^2{\bf r}\,e^{i{\bf qx}}\psi_{\bf k}^*({\bf r})\psi_{{\bf k}'}({\bf r})\\
&=&\sum_{\bf G} \delta({\bf q}-({\bf k}-{\bf k}')+{\bf G})\times\\
 &\,&\qquad\sum_{{\bf R}} e^{i{\bf k}'{\bf R}}
   \int d^2{\bf r}\,e^{i{\bf qr}}\phi^*({\bf r})\phi({\bf r}-{\bf R}).
\end{eqnarray*}
If we assume that the relevant electronic wavefunction is well localized,
we can ignore terms involving the overlap across different
sites (${\bf R} \neq 0$) in the last integral.  Then,
the only dependence of $J$ on {\bf k} and ${\bf k}'$ is through
the crystal momentum conservation condition, and
we find 
\begin{eqnarray}
\rho^{\rm phys}_{\bf q}(\epsilon)= S_{\bf q}\, \rho^{\rm lattice}_{\bf q}(\epsilon)
\label{rho_exp}
\end{eqnarray}
with
\begin{eqnarray*}
\rho^{\rm lattice}_{\bf q}(\epsilon) &=& 
 \sum_{{\bf k},n}\delta(\epsilon-E_n) 
 a_{\bf k}^{*n}a_{{\bf k}+{\bf q}}^{n} 
\nonumber\\
S_{\bf q} &=&
\int d^2{\bf r}\, |\phi({\bf r})|^2 e^{i{\bf qr}}.
\end{eqnarray*}
One immediately recognizes that in (\ref{rho_exp})
$\rho^{\rm lattice}_{\bf q}(\epsilon)$
is the Fourier component of the lattice density of states
that we analyzed in the earlier sections,
and $S_{\bf q}$ is the structure factor determined
by the atomic wavefunctions.
Peaks in the STM spectra arise from $\rho^{\rm lattice}_{\bf q}(\epsilon)$,
whereas $S_{\bf q}$ only provides  additional wavevector dependence.
Hence in our tight binding model we expect that wavevectors which differ
only by reciprocal lattice vectors have peaks
at the same energies, but with generally
different intensities.

\subsection{Current Normalization Condition}

An additional subtlety of STM experiments in
\cite{davis1,kapitulnik1,davis2,davis3,kapitulnik2} is the
space-dependent normalization used.
It is natural to assume that the tunneling matrix elements do not
change appreciably with energy over the energy range of interest.
Thus, if $z$ is the height of the STM tip above the sample, and
${\bf r}$ is its 2D coordinate along the plane of the sample surface,
then the differential tunneling conductance $g$ can be written as
\begin{eqnarray*}
g({\bf r},z,\epsilon)=f({\bf r},z)\rho^{\rm phys}({\bf r},\epsilon),
\end{eqnarray*}
where $\rho^{\rm phys}({\bf r},\epsilon)$ is the 2D density of states
in the CuO plane.  The experiments in
\cite{davis1,kapitulnik1,davis2,davis3,kapitulnik2} adjust the $z$
coordinate at every point ${\bf r}$ along the surface, so as to
keep the current at $V_f$ fixed at a predetermined value $I_f$. The
differential conductance normalized in this fashion is
\begin{eqnarray*}
g^{\rm meas}({\bf r},\epsilon)=f({\bf r})\rho^{\rm phys}({\bf r},\epsilon),
\end{eqnarray*}
where $f({\bf r})=I_f/\int_0^{eV_f}d\epsilon\,\rho^{\rm phys}({\bf
r},\epsilon)$.  Let us now discuss some properties of
$g^{\rm meas}({\bf r},\epsilon)$.

The spatial variation in $f({\bf r})$ is dominated by
the inhomogeneous quasiparticle weight within a unit cell.  To see
this, write
\begin{eqnarray*}
\rho^{\rm phys}({\bf r},\epsilon)=\rho_{\rm per}({\bf r},\epsilon)+\rho_{\rm TSB}({\bf r},\epsilon),
\end{eqnarray*}
where $\rho_{\rm per}({\bf r},\epsilon)$ is periodic with the lattice and is of order
one, whereas $\rho_{\rm TSB}({\bf r},\epsilon)$ breaks lattice translational symmetry and
is of order $V_0$ in our formalism.  If we define 
$f_{\rm per}({\bf r})=I_f/\int_0^{eV_f}d\epsilon\,\rho_{\rm per}({\bf r},\epsilon),$
then
\begin{eqnarray*}
f({\bf r})=f_{\rm per}({\bf r})(1-H({\bf r}))
\end{eqnarray*}
in terms of a TSB function $H({\bf r})$ of order $V_0$.

It is convenient to absorb $f_{\rm per}$ into $\rho$ by introducing a new function, 
$\rho'({\bf r},\epsilon)\equiv f_{\rm per}({\bf r})\rho^{\rm phys}({\bf r},\epsilon)$.
Due to the symmetry properties of $f_{\rm per}$, $\rho'$ is simply related to 
$\rho^{\rm lattice}$ through a modified structure factor 
\begin{eqnarray}
\rho'_{\bf q}(\epsilon)&=& S'_{\bf q}\,\rho^{\rm lattice}_{\bf q}(\epsilon)
\nonumber\\
S'_{\bf q} &=&
\int d^2{\bf r}\,  f_{\rm per}({\bf r}) |\phi({\bf r})|^2 e^{i{\bf qr}}.
\nonumber
\end{eqnarray}
Expressing $g^{\rm meas}$ in terms of $\rho'$,
\begin{eqnarray}
g^{\rm meas}_{\bf q}(\epsilon)=\rho'_{\bf q}(\epsilon)-
  \int\frac{d^2 {\bf k}}{(2\pi)^2}H_{{\bf q}-{\bf k}}\,\rho'_{\bf k}(\epsilon),
\label{gmeas}
\end{eqnarray}
we see that $g^{\rm meas}_{\bf q}(\epsilon)$ gets ``direct'' contributions
from structure in the LDOS at wavevector ${\bf q}$, as well as ``shadow''
contributions from structure in the LDOS at other wavevectors
${\bf k}$, whenever $H_{{\bf q}-{\bf k}}$ is non-zero.  Whereas (\ref{gmeas})
is an exact relation, it is useful to truncate it to order $V_0$ by keeping
in the second term only those contributions coming from the neighborhood of
the reciprocal vectors ${\bf k}\approx{\bf G}$,
\begin{eqnarray}
g^{\rm meas}_{\bf q}(\epsilon)&=&
\rho'_{\bf q}(\epsilon)-\alpha_{\bf q}\,\rho^{\rm lattice}_{{\bf k}=0}(\epsilon)
+O(V_0^2)
\label{gapprox0}\\
\alpha_{\bf q} &=& \sum_G H_{{\bf q}-{\bf G}}\,S'_{\bf G} 
\label{gapprox}
\end{eqnarray}
In this approximation
the shadow contribution to $g^{\rm meas}_{\bf q}(\epsilon)$
factorizes into the space-dependent factor $\alpha_{\bf q}$
and the space-averaged density of states
$\rho^{\rm lattice}_{{\bf k}=0} \propto 
\langle g^{\rm meas}(\epsilon) \rangle$. 
From (\ref{gapprox0}) and
(\ref{gapprox}) we can verify an important
property of the tunneling spectra
\begin{eqnarray}
\frac{g^{\rm meas}_{{\bf q}+{\bf G}}(\epsilon)}{g^{\rm meas}_{\bf q}(\epsilon)}
=\frac{S'_{{\bf q}+{\bf G}}}{S'_{\bf q}}
\nonumber
\end{eqnarray}
when ${\bf G}$ is a vector of the reciprocal lattice and ${\bf q}$ is not.
Hence, we expect  $g^{\rm meas}_{\bf q}(\epsilon)$
and $g^{\rm meas}_{{\bf q}+{\bf G}}(\epsilon)$
to have peaks at the same energies
but in general with different overall amplitudes.


An interesting question to ask is whether
it is possible to analyze experimental data
in a way that would allow to separate direct and shadow
contributions to $g^{\rm meas}_{\bf q}(\epsilon)$.
Below we demonstrate that this is possible using
an exact sum rule obeyed by $\rho^{\rm lattice}_{\bf q}(\epsilon)$.
Regardless of the model used and the
nature of the symmetry breaking perturbation, the sum over all frequencies of
$\rho^{\rm lattice}_{\bf q}(\epsilon)$ should be identically zero
for all ${\bf q}$ different from the reciprocal lattice vectors
${\bf G}$:
\begin{eqnarray*}
\int_{-\infty}^\infty d\epsilon\,\rho^{\rm lattice}_{\bf q}(\epsilon)&=&
\frac{1}{N}\sum_{{\bf k}\sigma} \langle 0|\left\{
c_{{\bf k}+{\bf q}\sigma},c_{{\bf k}\sigma}^\dagger
\right\}|0\rangle\\
&\equiv& 2\sum_{\bf G}(2\pi)^2\delta({\bf q}-{\bf G}).
\end{eqnarray*}
In principle, this identity can be used to remove the shadow contribution
in equation (\ref{gmeas}).  In particular, for ${\bf q}\ne{\bf G}$,
 combining the approximate result
(\ref{gapprox0}) and our knowledge of $\rho^{\rm lattice}_{{\bf k}=0}(\epsilon)$
from experiment, we can fix $\alpha_{\bf q}$ by requiring the sum rule to be
obeyed:
\begin{eqnarray}
\rho'_{\bf q}(\epsilon) &=&
g^{\rm meas}_{\bf q}(\epsilon) 
- \langle g^{\rm meas}(\epsilon)\rangle
\frac{\int_{-V_{\rm MAX}}^{V_{\rm MAX}}g^{\rm meas}_{\bf q}(\epsilon) d \epsilon}
{\int_{-V_{\rm MAX}}^{V_{\rm MAX}}\langle g^{\rm meas}(\epsilon)\rangle d \epsilon}
\label{shadowremoved}
\end{eqnarray}
Here $V_{\rm MAX}$ should be chosen sufficiently large so
that the ratio of the two integrals is close to its
saturated value, yet it should be small enough that we are 
still justified in using a single band model
and a local picture of electron tunneling.

For completeness, we also list two other
sum rules obeyed by tunneling spectra.
By construction, at every wavevector ${\bf q}\ne 0$
the function $g^{\rm meas}_{\bf q}(\epsilon)$ must satisfy
the normalization condition
\begin{eqnarray*}
\int_{0}^{eV_f} d\epsilon\,g^{\rm meas}_{\bf q}(\epsilon)= 0.
\end{eqnarray*}

One can derive an independent sum rule if we restrict the class of symmetry
breaking Hamiltonians to effective one-particle operators (\ref{general})
[This includes all perturbations considered in this work.] 
Then the $\epsilon$-weighted average of $\rho^{\rm lattice}_{\bf q}(\epsilon)$
will be, for ${\bf q}\ne{\bf G}$,
\begin{eqnarray*}
\int_{-\infty}^\infty d\epsilon\,\epsilon\,\rho^{\rm lattice}_{\bf q}(\epsilon)
&=&\frac{1}{N} \sum_{{\bf k}\sigma}
\langle 0 |\left\{ \left[c_{{\bf k}+{\bf q}\sigma},H\right],
c^\dagger_{{\bf k}\sigma}\right\} | 0 \rangle\\
&=&\frac{2}{N} \sum_{{\bf k}\sigma}V_{{\bf kq}}.
\end{eqnarray*}
For the basis functions discussed in Section II we find
that only the $V_{\bf kq}$ describing simple CDW
gives finite contributions after summing over ${\bf k}$.  Hence,
\begin{eqnarray*}
\int_{-\infty}^\infty d\epsilon\,\epsilon\,\rho^{\rm lattice}_{\bf q}(\epsilon)
=2 V_{\bf q}^{\rm cdw}.
\end{eqnarray*}
It is important to point
out that this sum rule will be spoiled by shadow contributions in (\ref{gmeas}),
and is only of use if these have been previously removed, using
for example procedure in equation (\ref{shadowremoved}).

As a useful consistency check of our formalism, one can easily verify
that expressions (\ref{lincomb}), (\ref{Afinal}), and (\ref{Bfinal}) satisfy both
sum rules.  We emphasize, however, that although these expressions
are only correct to linear order in perturbation strength,
the sum rules are non-perturbative and therefore hold to all orders
in perturbation theory.
Furthermore, their validity is not affected by the
introduction of finite quasiparticle
lifetimes as, for any normalized symmetric distribution $g(\epsilon)$,
$\int d\epsilon\, (\alpha+\beta\epsilon) g(\epsilon-\epsilon_0)=
\alpha+\beta\epsilon_0$.  By contrast, the average of $\rho_{\bf q}(\epsilon)$
weighted by any other power of $\epsilon$ is sensitive to details of
quasiparticle smearing.

Unfortunately, these sum rules are of limited
immediate use, since the bulk of the integration comes from large energies,
whereas current experiments only probe a relatively narrow range of biases
about the chemical potential.

\section{Photoemission}


Before concluding, we would like to propose
a way of identifying weak charge ordering
in photoemission experiments that could supplement
current STM studies.  A common
signature of a strong charge ordering in the ARPES experiments is the
presence of shadow bands: the electron spectral function at momentum
${\bf k}$ acquires an additional peak at the energy of the
quasiparticle at momentum ${\bf k}+{\bf Q}$.  For weak charge order
the shadow bands may be difficult to observe: when the energy
difference between $E_{\bf k}$ and $E_{{\bf k}+{\bf Q}}$ is large,
mixing between quasiparticles is negligible and the intensity of the
shadow peaks is vanishingly small. Strong mixing only occurs when
states ${\bf k}$ and ${\bf k}+{\bf Q}$ are nearly degenerate, although
in this case the two peaks are hard to distinguish since they are
close in energy. Thus we expect to observe an increase of the apparent
linewidth of quasiparticles when the latter satisfy the degeneracy
condition $E_{\bf k} = E_{{\bf k}+{\bf Q}}$ and are strongly affected
by the charge order.  For example, in the case of the
\BSCCO band
structure shown on Figure \ref{fig4} we expect an anomalous increase in the
apparent quasiparticle linewidth at the points A, A', B, and B' on
the Fermi surface.

\section{Conclusions}

To summarize, we considered the effects of 
weak translational symmetry breaking on the $d$-wave
superconducting state of the cuprates. For systems with
periodic charge order we derived an explicit
formula for the energy dependence
of the Fourier component of the local density of states
for several types of order, including simple charge density wave,
electron kinetic energy and superconducting gap
modulations. We argued that within a one band model
the STM spectra observed in 
\cite{davis1,kapitulnik1,kapitulnik2} cannot be explained by a simple
charge density wave but require the existence
of some form of (anomalous) dimerization,
i.e. modulation in the electron hopping or
in the superconducting pairing amplitude.  
We discussed a situation in which charge order
has finite correlation length due to pinning
by impurities. In this case the LDOS has Fourier components
for a range of momenta around the ordering wavevector
${\bf Q}$. For different wavevectors ${\bf q}$,
peaks in $\rho_{\bf q}(\epsilon)$ will occur at
different energies, although the peak dispersion is weak,
in agreement with \cite{davis2,kapitulnik2}.
We also considered systems in which translational
symmetry breaking  comes not from charge ordering but from
impurities. We found that the Fourier components of the LDOS
in this case have peaks for a wide range of wavevectors and strong
dispersion of these peaks is consistent with the
STM experiments of Refs. \cite{davis2,davis3}.

We thank J.C. Campuzano, J.C. Davis,  J.P. Hu, 
A. Kaminski, A. Kapitulnik, S. Kivelson, S. Sachdev,
and S.C. Zhang for illuminating discussions.
This work was supported by the Nanoscale Science
and Engineering Initiative of the National Science Foundation
under NSF Award Number PHY-0117795 and by 
NSF grants  DMR-9981283, 
DMR-9714725, and DMR-9976621

\newpage

\begin{figure}
\epsfxsize=10cm \centerline{\epsffile{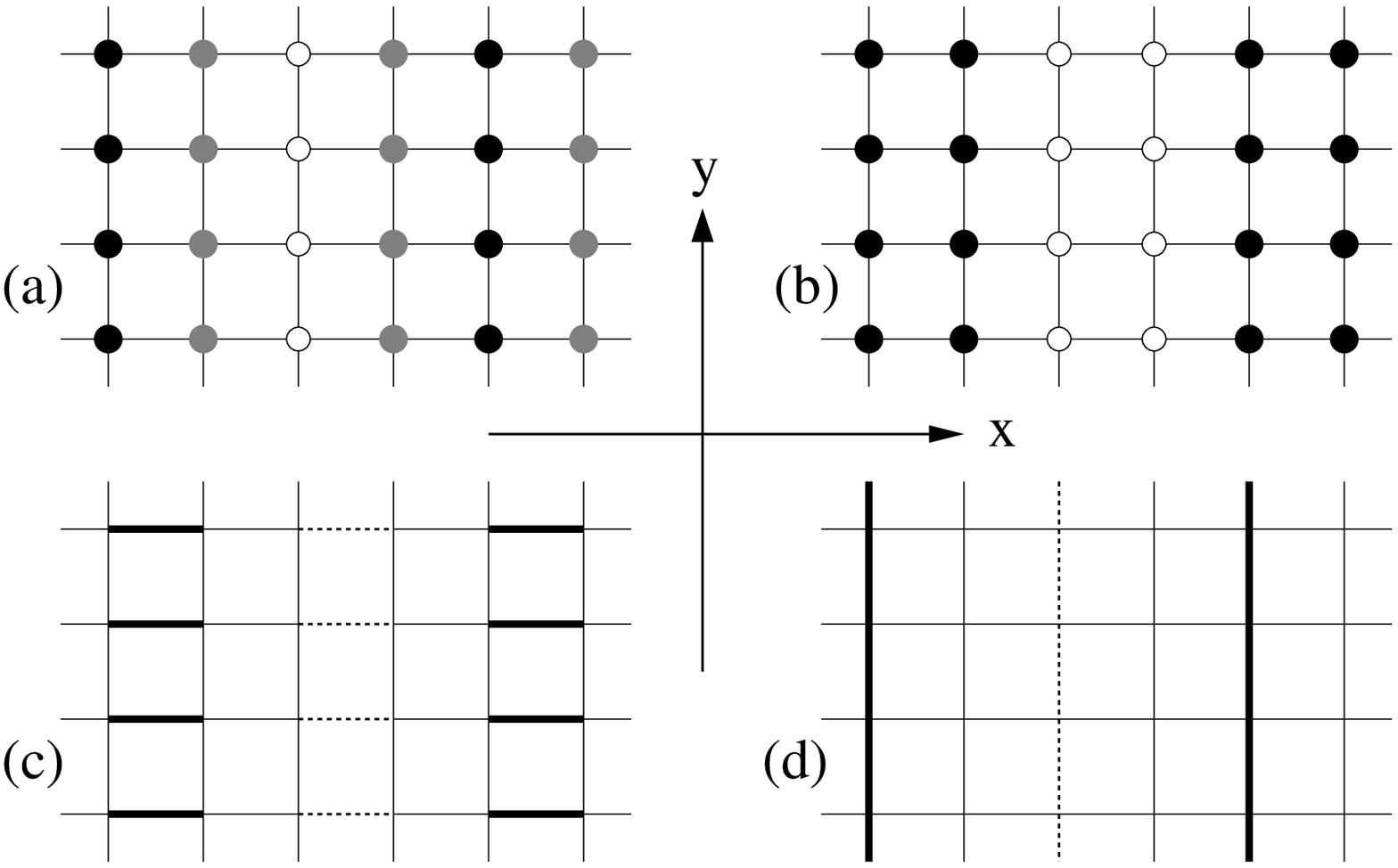}} \caption{ Order
parameters with wavevector ${\bf Q}=(\frac{2\pi}{4},0)$ considered in
this paper. Figures (a) and (b) correspond to site and bond centered CDW
respectively. Black circles correspond to sites of higher electron
density, white circles to sites of lower electron density, and gray
circles to sites with the average electron density. 
Figures (c) and (d) describe longitudinal and transverse
dimerizations respectively.  Heavy lines correspond to bonds
with higher tunneling amplitude, and dotted lines
to bonds with lower tunneling amplitude.
Anomalous dimerization may be shown schematically as
on figures (c) and (d), with heavy and dotted bonds describing
higher and lower pairing amplitudes.
 } \label{fig1}
\end{figure}

\begin{figure}
\epsfxsize=10cm \centerline{\epsffile{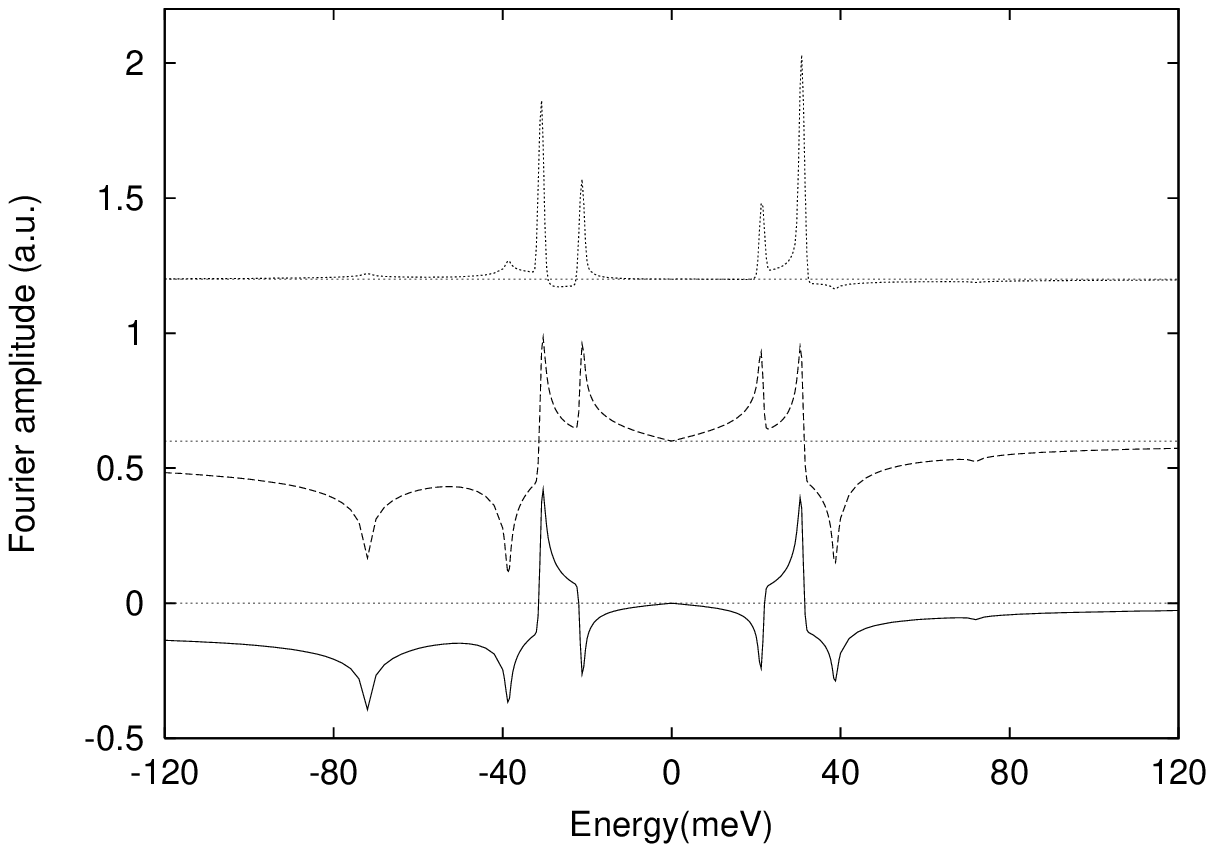}} \caption{
Energy dependence of the Fourier component of the local
density of states $\rho_{\bf Q}(\epsilon)$ at ${\bf Q}=(2\pi/4,0)$
for various cases of charge ordering.
\BSCCO type band structure is assumed.
We show a direct calculation based on formulas 
(\ref{lincomb},\ref{Afinal},\ref{Bfinal}).
The curves correspond to the 
CDW (solid), longitudinal dimerization (dashed) and
anomalous longitudinal dimerization (dotted) orders. 
To simplify the comparison,  
$\rho_{\bf Q}(\epsilon)$ is multiplied by $-1$
for CDW, and by $\frac{1}{2}$ for anomalous longitudinal
dimerization. In addition, subsequent curves are shifted vertically
by 0.6.  Results for both kinds of transverse dimerization are 
qualitatively similar to corresponding longitudinal results
and are omitted for visual clarity.
} \label{fig2}
\end{figure}

\begin{figure}
\epsfxsize=10cm \centerline{\epsffile{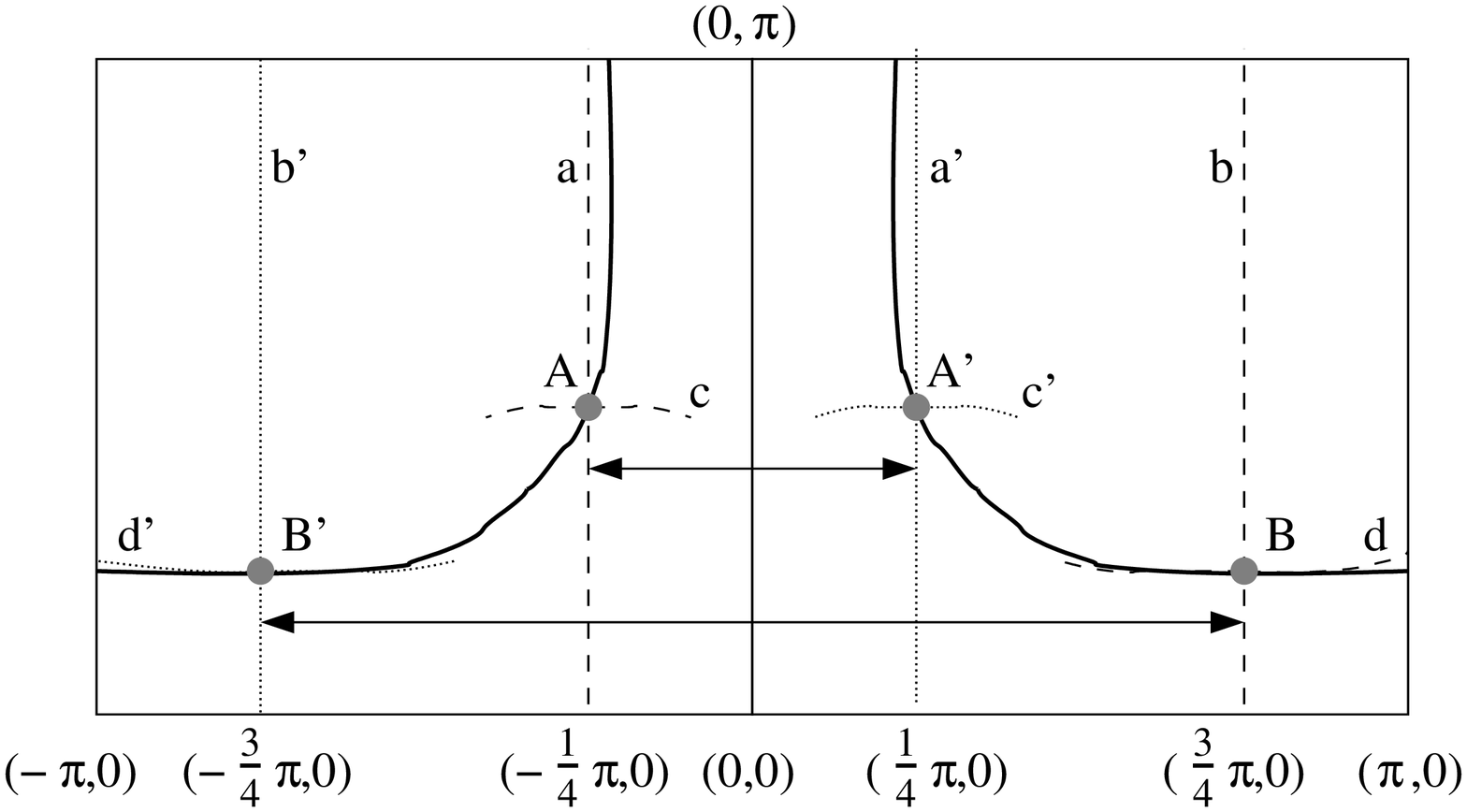}} \caption{ 
Fermi surface for \BSCCO.  Dashed lines correspond to the
quasiparticles that satisfy $E_{{\bf k} + {\bf Q}}=E_{\bf k}$ and are
strongly affected by $\Delta {\cal H}$ when ${\bf Q}=(2\pi/4,0)$.
Quasiparticles on line $a$ hybridize with quasiparticles on line $a'$
(and similarly for lines $b$ and $b'$, and curves $c,$ $c'$ and $d,$
$d'$).  Crossings of these lines with the Fermi surface (points A and
A', B and B') give the minimal energy of such quasiparticles:
$0.5 \Delta_0$ and $0.7 \Delta_0$ respectively.
Contributions from these points produce sharp peaks at energies
$20$ and $30$ meV in Fig. 2.  The Van Hove
singularity for the Bogoliubov quasiparticles at energy $\Delta_0$
leads to a peak at $40$ meV.  } 
\label{fig4}
\end{figure}

\begin{figure}
\epsfxsize=10cm \centerline{\epsffile{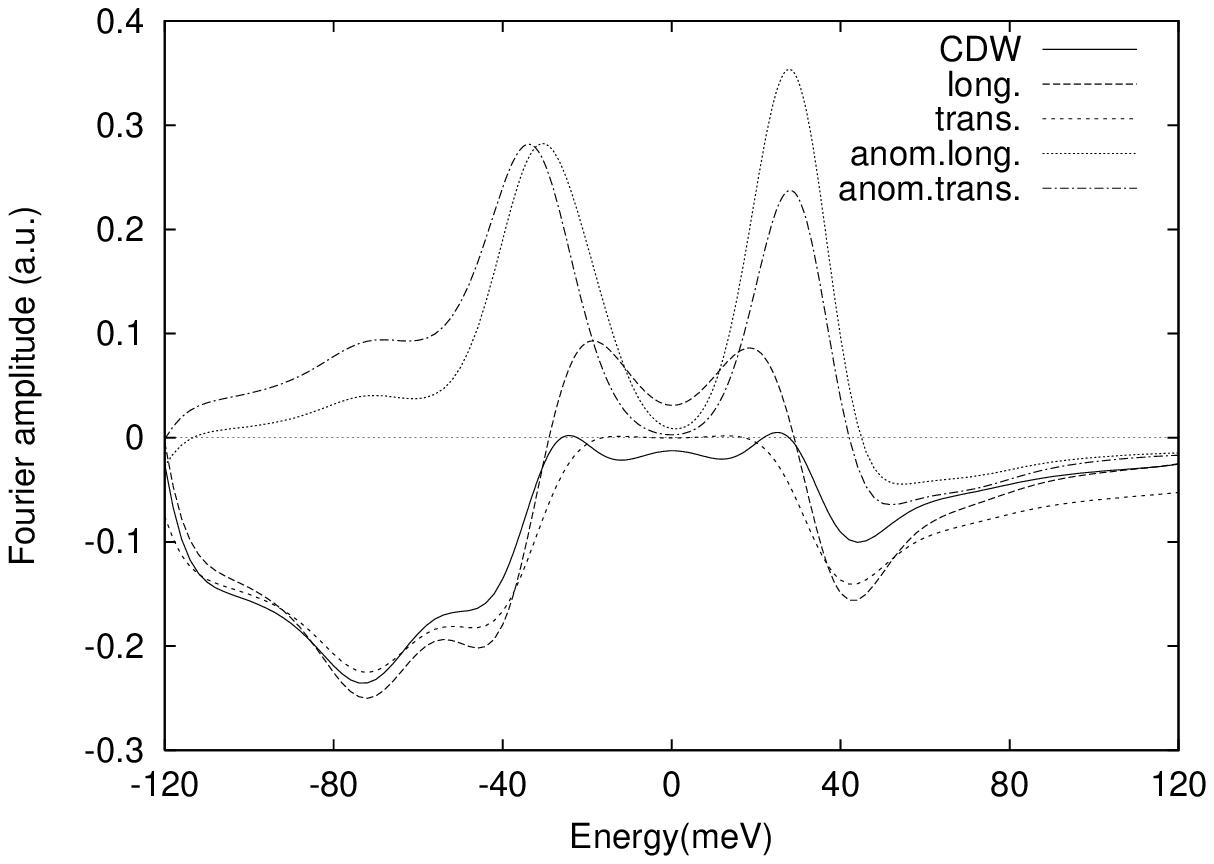}} \caption{
The results in Fig.~\ref{fig2} are shown after
smearing over an energy range of 8 meV.
To simplify the comparison,  
$\rho_{\bf Q}(\epsilon)$ was multiplied by $-1$
for CDW and anomalous transverse dimerization,
and by a factor of $-\frac{1}{2}$ for transverse
dimerization.
} \label{fig3}
\end{figure}

\begin{figure}
\epsfxsize=10cm \centerline{\epsffile{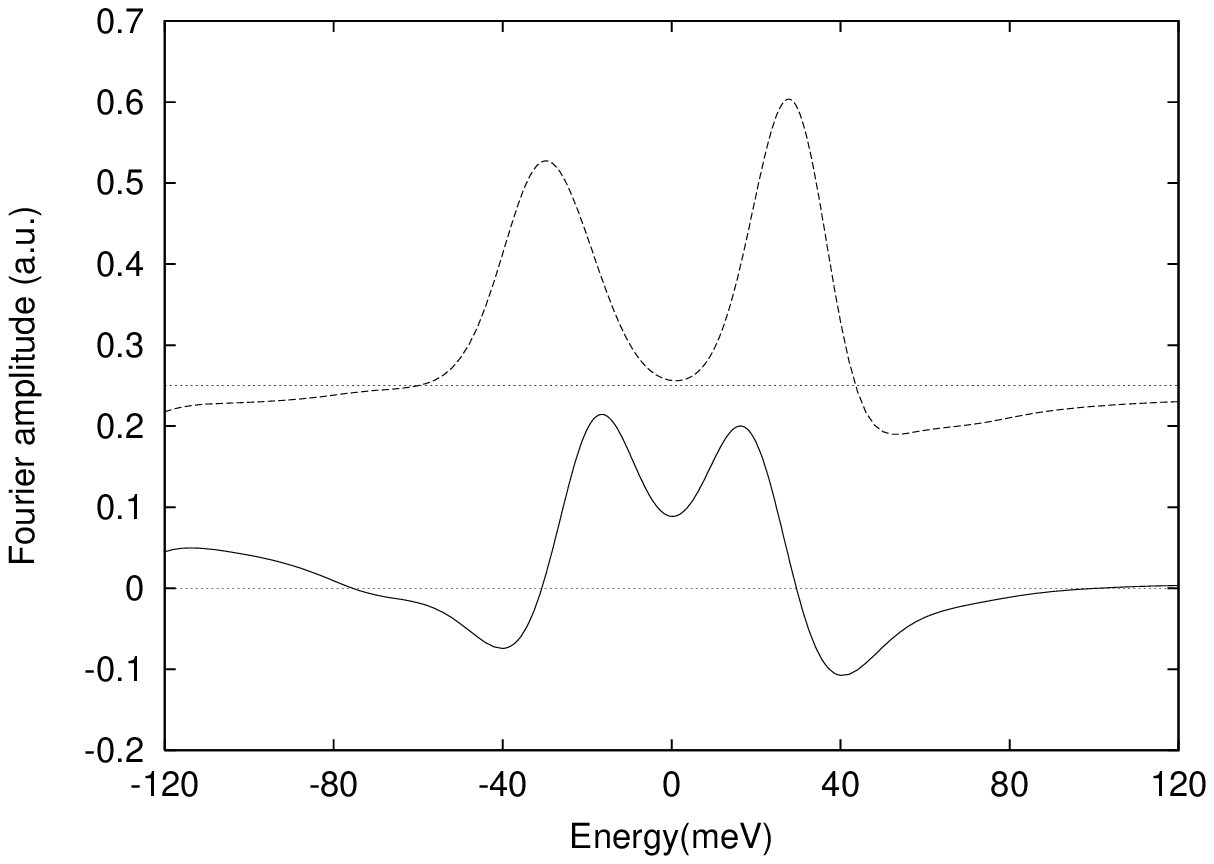}} \caption{ 
$\rho_{Q}(\epsilon)$ for \BSCCO type band structure,
ordering wavevector $Q=2\pi/4$ and 
a combination of charge orders: longitudinal dimerization and CDW,
$1.05({\rm CDW})+({\rm long.dim.})$
(solid line);
anomalous longitudinal dimerization and CDW,
 $({\rm anom.long.dim.})+0.2({\rm CDW})$
(long-dashed line). The same smearing is assumed
as in Fig.\ref{fig3}.
For clarity, curves
have been offset vertically by 0.25.
}\label{fig5}
\end{figure}

\begin{figure}
\epsfxsize=10cm \centerline{\epsffile{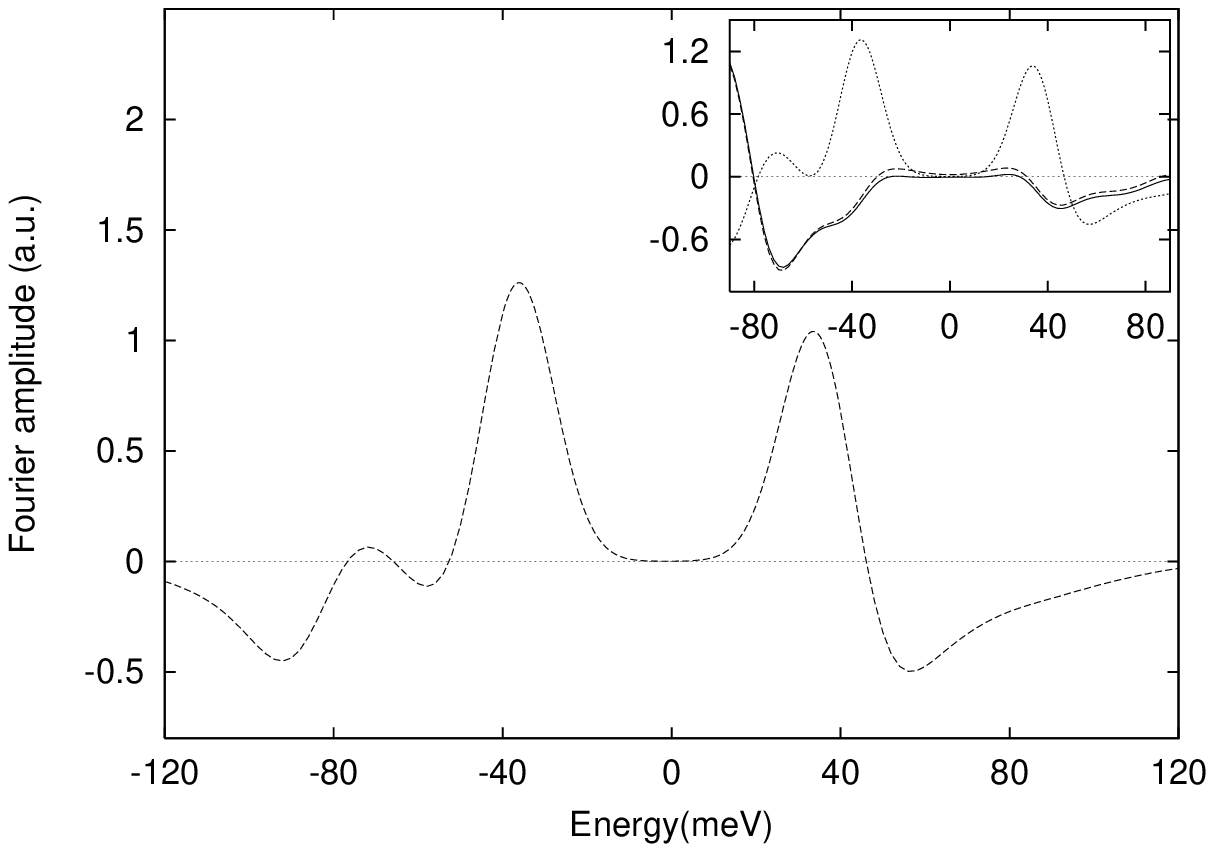}} \caption{
\YBCO type band structure
and ordering wavevector ${\bf Q}=(2\pi/8,0)$. The inset
shows $\rho_{Q}(\epsilon)$ for CDW (solid line), dimerization
(dashed line), and anomalous
dimerization (dotted line) separately. Main figure
has the linear combination $({\rm ald})+0.2({\rm CDW})$,
which was displayed for \BSCCO in Fig.~\ref{fig5} (the other
linear combination is nearly zero and thus omitted). The same smearing 
is assumed as in Fig.~\ref{fig3}.
}\label{fig6}
\end{figure}

\begin{figure}
\epsfxsize=10cm \centerline{\epsffile{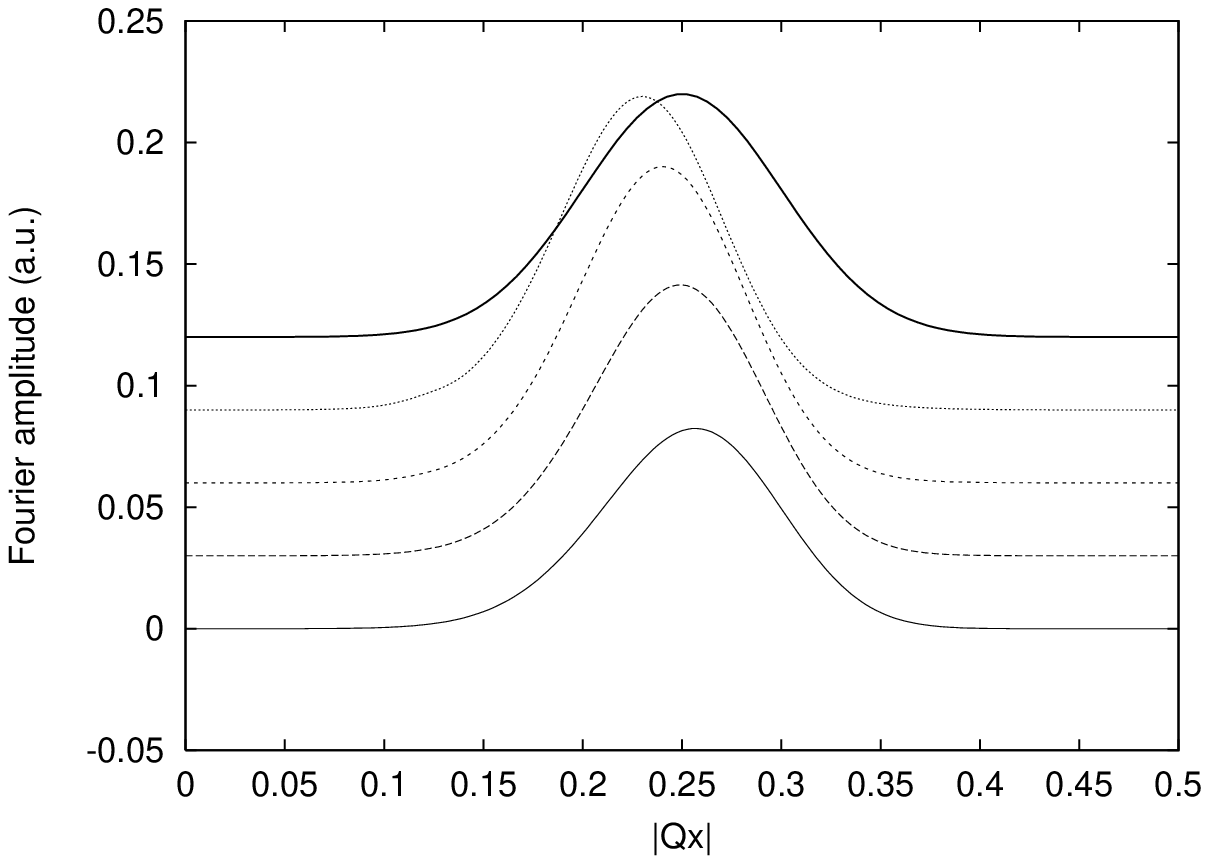}} \caption{
Dispersion in the $(0,0)$ to $(\pi,0)$ direction 
in a system with charge order with randomness
(momentum is measured in units of $2\pi$).
Charge order is assumed to have Gaussian distribution centered
around wavevector $(2\pi/4,0)$ with the width $2\pi/20$.
The function $V({\bf q})$ in (\ref{specific})
is shown, up to a scale, as the thick solid curve.
For visual clarity, only results corresponding
to the linear combination $1.05({\rm CDW})+({\rm ld})$ are displayed.  Each
curve corresponds to a different bias; starting from the bottom,
the biases are 8 mV, 12 mV, 16 mV, and 20 mV.
Throughout, the quasiparticle smearing is fixed at 8 meV.
}\label{fig7}
\end{figure}

\begin{figure}
\epsfxsize=10cm \centerline{\epsffile{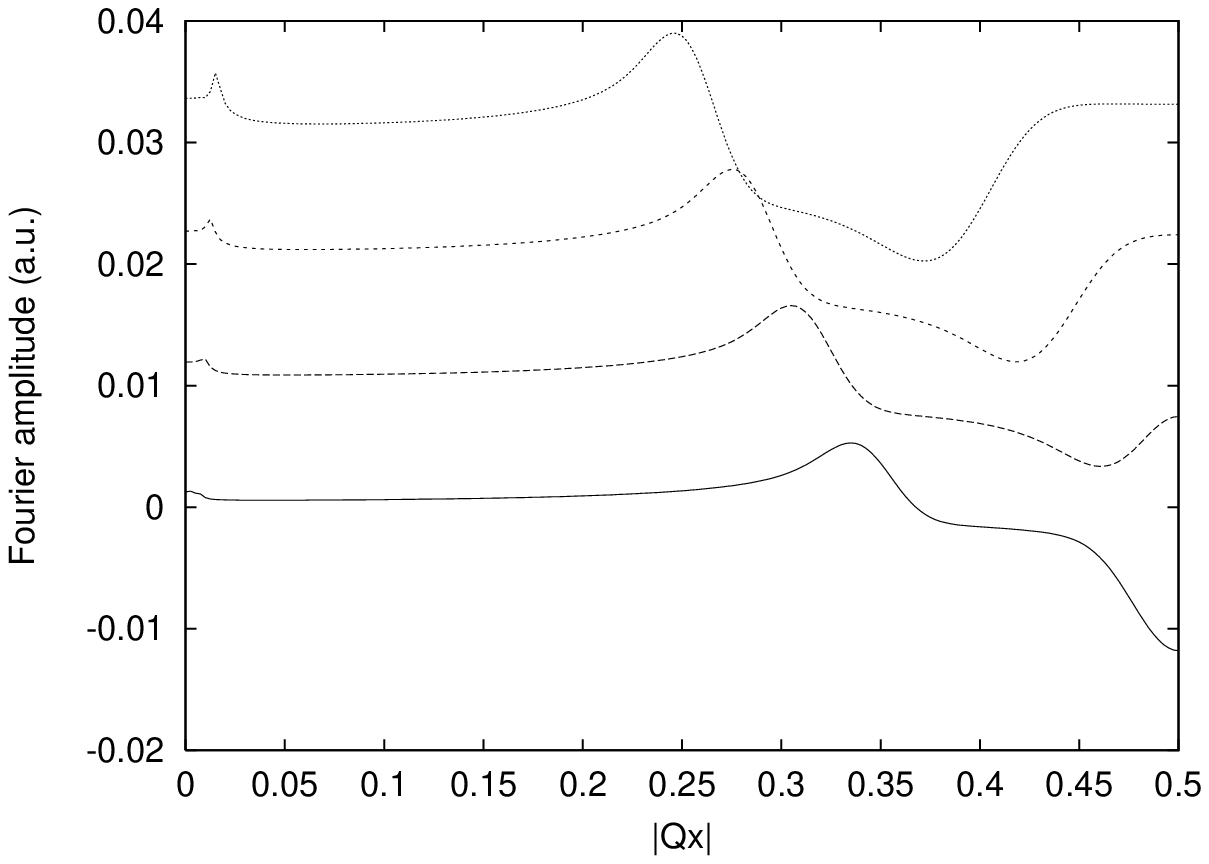}} \caption{
Dispersion in the $(0,0)$ to $(\pi,0)$ direction 
in a model with impurity induced quasiparticle
scattering (momentum is measured in units of $2\pi$).
Each curve corresponds to a different bias; starting from the bottom,
the biases are 8 mV, 12 mV, 16 mV, and 20 mV.
Unlike other computations in this paper,
the quasiparticle smearing is fixed at 2 meV.  This is done since 
the main features in these curves  are averaged out for the
usual smearing of 8 meV. 
}\label{fig8}
\end{figure}

\begin{figure}
\epsfxsize=10cm \centerline{\epsffile{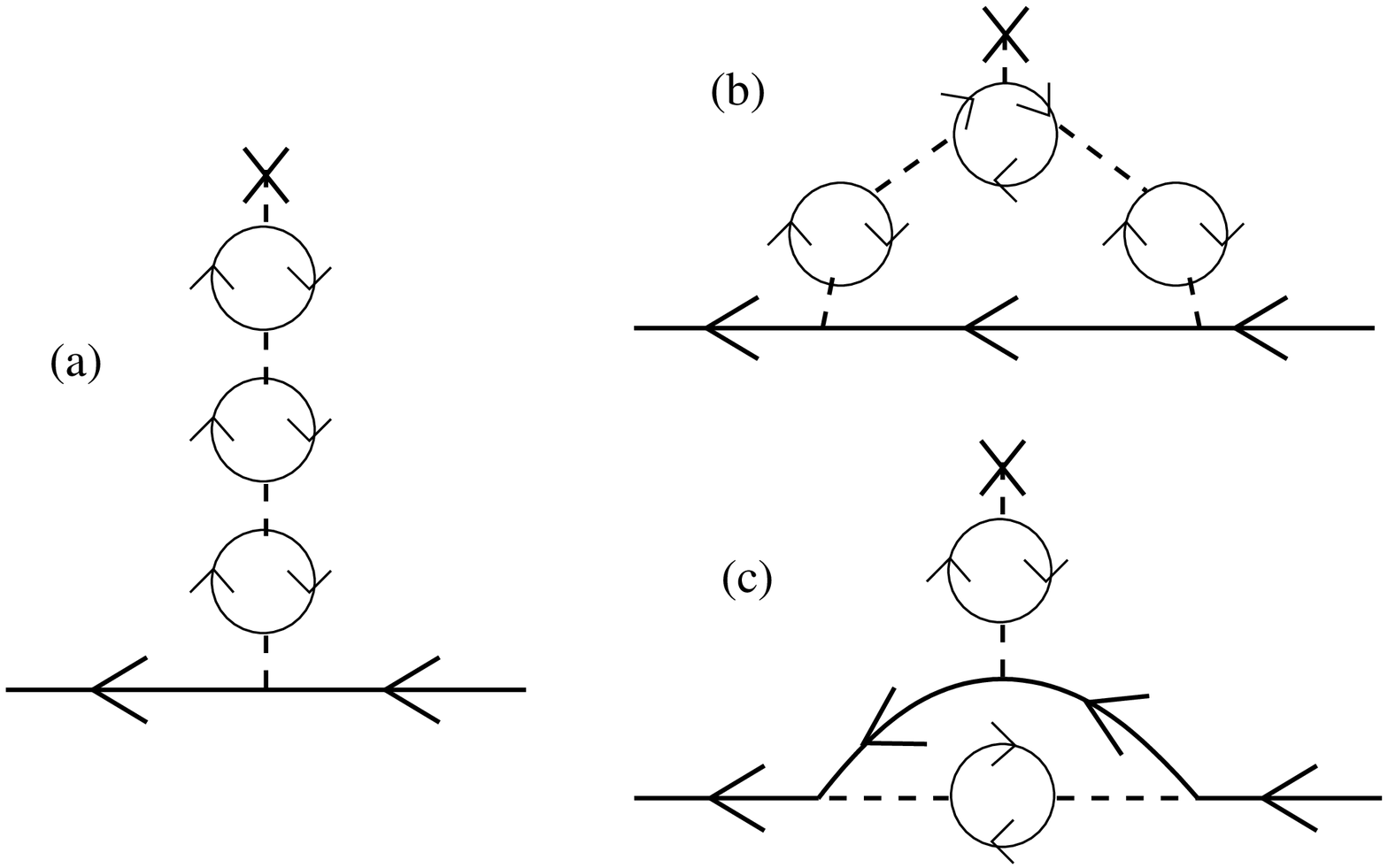}} \caption{
{\bf (a)} Diagram of Hartree type used in computing the RPA response
for a system of interacting electrons in the vicinity of a CDW
instability.  An external field (e.g. an impurity potential) pins the CDW.
{\bf (b)} and {\bf (c)}~Contributions beyond the Hartree approximation, which may be subdominant
depending on the model used.  Their inclusion leads to a frequency-dependent
self energy. 
}\label{fig9}
\end{figure}


\begin{thebibliography}{}
\bibitem{note}
The distinction between site and bond-centered orders
is only defined for translational symmetry breaking commensurate
with the lattice, i.e. when the wavevector is $Q=2\pi/n$ with $n$
integer.
\end{thebibliography}
\end{document}